# ANALYSIS OF LUMINESCENCE SPECTRA OF SUBSTRATE-FREE ICOSAHEDRAL AND CRYSTALLINE CLUSTERS OF ARGON

Yu.S. Doronin, V.L. Vakula, G.V. Kamarchuk, A.A. Tkachenko, V.N. Samovarov

*B. Verkin Institute for Low Temperature Physics and Engineering of the NAS of Ukraine*
*47 Nauky Ave., Kharkiv, 61103, Ukraine*
E-mail: vakula@ilt.kharkov.ua

**Abstract**

We propose a new approach to analysis of cathodoluminescence spectra of substrate-free nanoclusters of argon produced in a supersonic jet expanding into a vacuum. It is employed to analyze intensities of the luminescence bands of neutral and charged excimer complexes $(Ar_2)^*$ and $(Ar_4^+)^*$ measured for clusters with an average size of 500 to 8900 atoms per cluster and diameters ranging from 32 to 87 Å. Concentration of the jet substance condensed into clusters, which determines the absolute values of the integrated band intensities, is shown to be proportional to the logarithm of the average cluster size. Analysis of reduced intensities of the $(Ar_2)^*$ and $(Ar_4^+)^*$ bands in the spectra of crystalline clusters with an fcc structure allows us to conclude that emission of the neutral molecules $(Ar_2)^*$ comes from within the whole volume of the cluster, while the charged complexes $(Ar_4^+)^*$ radiate from its near-surface layers. We find the cluster size range in which the jet is dominated by quasicrystalline clusters with an icosahedral structure and demonstrate that the transition from icosahedral to fcc clusters occurs when the average cluster size is $\overline{N}$ = 1400±400 atoms per cluster.

**Introduction**

Nanosized aggregates of atoms and molecules, called nanoclusters, are an intermediate between single atoms and macroscopic bodies. The fundamental interest in these systems has largely been due to a number of unusual properties they display which are not observed in bulk samples. Among them are, for example, dependence of cluster structure on cluster size and presence in clusters of stable quasicrystalline structures with a 5-fold axis of symmetry [1,2]. Van der Waals clusters of solid rare gases, such as, for example, argon, can serve as good model objects. The powerful experimental methods for studying nanosystems include spectroscopic techniques like luminescence study of substrate-free clusters generated in supersonic jets (see, e.g., [3,4]). One of the main difficulties of spectroscopic studies is to correctly take into account the composition of the supersonic jet and its influence on the characteristics of the resulting spectra such as integrated intensity of emission bands. The purpose of the present paper is to develop a method which would make it possible to use integrated intensities to analyze the various processes occurring in clusters produced in supersonic jets, including transformations of the cluster structure.

**Experimental**

We present our cathodoluminescence study of substrate-free solid clusters of argon formed in the process of condensation of argon gas in a supersonic jet which expands adiabatically into a vacuum through a conical nozzle. Cluster size and structure were changed by varying the gas temperature at the nozzle inlet, the stagnation pressure $p_0$ was kept equal to 1 atm. A decrease in the gas temperature $T_0$ resulted in bigger clusters. The average cluster size was calculated using a modified Hagena relation obtained in electron-diffraction studies performed with a nozzle which was analogous to ours [5].

The clusters studied in the present paper had an average size ranging from 500 to 8900 atoms per cluster (at/cl) and a diameter of 32-87 Å, which corresponded to the temperature range of the gas at the nozzle inlet $T_0$ = 116-240 K. The cluster temperature was about 40 K and virtually independent of the gas parameters at the nozzle inlet [6].

To obtain cathodoluminescence spectra, clusters were excited by a 1-keV electron beam. The signal was registered in the photon energy range 8.1-11.8 eV containing emission bands of neutral and charged excimer complexes $(Ar_2)^*$ and $(Ar_4^+)^*$. For experimental technique details see Refs. [7,8].

**Results and discussion**

Figure 1 shows two examples of cathodoluminescence spectra for clusters with an average size $\overline{N} \approx 1000$ at/cl (cluster diameter being 42 Å) and 8900 at/cl (87 Å) and their decomposition into spectral components. The spectra measured for clusters of other sizes corresponded qualitatively with those in Fig. 1. The spectra consist of a few bands emitted from molecular and atomic states of argon. The narrow band at 11.61 eV corresponds to the $^3P_1 \rightarrow ^1S_0$ transition in single atoms of argon desorbed from clusters after their excitation by electrons and in argon atoms present in the gaseous fraction of the jet which did not condense into clusters. There are two pronounced molecular continua in the low-energy section of the spectra which are emitted from clusters: the band at 9.6 eV corresponds to transitions from vibrationally relaxed levels of neutral excimer complexes $(Ar_2)^*$ in the state $^3P_1+^1S_0$ and $^3P_2+^1S_0$ (see., e.g., [9]), while the band at 8.9 eV can be assigned to transitions in charged complexes $(Ar_4^+)^*$ [10,11]. In the higher-energy spectral range, we can see emission from partly vibrationally relaxed (W′ band at 10.6 eV) and vibrationally excited (asymmetric W band with a maximum at 11.3 eV) states of $(Ar_2)^*$ molecules in clusters. It should be noted that in bulk samples of solid argon the W band has a significantly lower relative intensity than it has in clusters [12,13]. Analysis of the W′ and W bands was only needed to fit more accurately the profiles of the molecular continua $(Ar_2)^*$ and $(Ar_4^+)^*$ which are of main interest in the present study.

Formation of neutral and charged excimer molecules comes as a result of self-trapping of free excitons whose lowest energy states are effectively filled upon irradiation of clusters with electrons. The self-trapping process is strongly intensified by a presence in the sample of impurities and various crystal lattice imperfections, which can also include the crystal surface. In pure bulk samples of solid rare gases with a perfect crystalline structure, observation of transitions from both coherent and localized excitonic states is possible at low temperatures. In the case of clusters, luminescence of free excitons has as yet been observed only for xenon with a small admixture of argon and explained by formation of exciton-impurity complexes [14]. Luminescence spectra from pure clusters of argon only display emission from localized states which comes mostly from neutral and charged excimer complexes $(Ar_2)^*$ and $(Ar_4^+)^*$. This emission is observed from both crystalline clusters and clusters with an icosahedral structure with no translational symmetry in the radial direction.

To analyze the physical processes of formation and relaxation of electronic excitations in clusters as function of their structure we need to study the evolution of the spectral features, their integrated intensity in the first place, upon a change in cluster size. Changing the average cluster size we also change the absolute and relative quantity of the jet substance condensed into clusters. This poses a problem of calculating integrated intensities of spectral bands normalized to one cluster of a given size. Till now, integrated intensities have been analyzed under the assumption that the fraction of the substance condensed into clusters depends weakly on the gas parameters at the nozzle inlet and, therefore, on the average size of clusters in the jet. To solve the problem it is convenient to deal with integrated intensities of the emission bands of vibrationally relaxed excimer complexes $(Ar_2)^*$ and $(Ar_4^+)^*$, since their emission comes almost exclusively from clusters and is strong enough to minimize the error in finding the intensity values.

Integrated intensity of the emission which corresponds to a certain radiative channel of relaxation of the electronic excitations depends, in the general case, on excitation cross-section for a unit volume of the emitting substance, density of the beam of exciting particles, probability of radiation occurring through a given radiative channel, and substance volume taking place in the excitation and radiation processes. In the case of clusters excited by an electron beam of constant density, it should depend on excitation cross-section for one cluster, number of excited clusters, and probability of relaxation of the produced excitations through the radiative channel under consideration:

$$I \sim \sigma_{exc} n_{cl} W_{rad}, \qquad (1)$$

where $I$ is integrated intensity of an emission band, $\sigma_{exc}$ is excitation cross-section for one cluster, $n_{cl}$ is number of excited clusters, and $W_{rad}$ is probability of radiation through the radiative channel which gives rise to the analyzed emission band.

The number of clusters $n_{cl}$ excited in the jet in unit time is given by the total number of atoms $n_0$ passing through the excitation area in unit time as well as by the fraction of substance $c_{cl}$ condensed into clusters and average cluster size $\overline{N}$ in the excitation area:

$$n_{cl} = c_{cl} n_0 / \overline{N}. \qquad (2)$$

It was shown theoretically in Ref. [15] that the maximum concentration $c_{max}$ of bound atoms at the end of expansion of pure atomic gas is

$$c_{max} \sim \frac{T_*}{\varepsilon_0} \ln N_{max}, \qquad (3)$$

where $N_{max}$ is the maximum number of cluster atoms at the end of the expansion process, $T_*$ is the condensation onset temperature, and $\varepsilon_0$ is the mean binding energy per cluster atom. In the case of big clusters, when the binding energy is weakly dependent on cluster size and is close to its value for a macroscopic system, the ratio $\frac{T_*}{\varepsilon_0}$ can be believed to be constant.

Taking into account that in our experiments the jet excitation area was 30 mm away from the nozzle outlet, that is where clusters already possess the largest average size possible and the process of their further growth is thus ineffective, and assuming that the maximum cluster size $N_{max}$ in the excitation area is proportional to the weight-average cluster size $\overline{N}$, we can conclude that the fraction of the substance $c_{cl}$ condensed into clusters is given by Eq. (3) with a numerical factor under the logarithm:

$$n_{cl} \sim n_0 \ln \overline{N} / \overline{N} + A\, n_0 / \overline{N}, \qquad (4)$$

where A is the logarithm of the numerical factor. Then we have

$$n_{cl} \sim n_0 \ln \overline{N} / \overline{N} \qquad (5)$$

under the assumption that the second term in Eq. (4) is small for clusters of the sizes studied in the paper. The value $n_0$ depends on the stagnation pressure $p_0$, which was kept constant in our experiments, so

$$n_{cl} \sim \ln \overline{N} / \overline{N}. \qquad (6)$$

Cross-section of cluster excitation by an electron is, in the general case, a function of the electron energy and cluster size. In the case of clusters excited by electrons of the same energy, we can use the cluster geometrical cross-section, which is proportional to $\overline{N}^{2/3}$, as its excitation cross-section. Therefore, Eq. (1) can be rewritten as follows:

$$I \sim \overline{N}^{2/3} n_{cl} W_{rad} \sim \overline{N}^{-1/3} \ln \overline{N} \; W_{rad}. \qquad (7)$$

Equation (7) allows us, by using the experimental data on the integrated intensity $I$ of a cluster luminescence band, to get information on the value of $W_{rad}$, which reflects the physics of the processes leading to the radiative transitions resulting in the band formation which occur in one cluster.

Figure 2 shows the integrated intensities of the $(Ar_2)^*$ and $(Ar_4^+)^*$ bands divided by $\ln \overline{N}$ as functions of the average cluster size $\overline{N}$. Each point is a value averaged over several spectral measurements. Two cluster-size ranges can be seen in Fig. 2 which are characterized by different behavior of the reduced intensities of both of our bands.

The big-cluster section ($\overline{N} \geq 1800$ at/cl) displays a nonlinear growth of $I / \ln \overline{N}$ with increasing $\overline{N}$, which can be approximated with good accuracy by $I / \ln \overline{N} \sim \overline{N}^\alpha$ with $\alpha = 2/3$ for $(Ar_2)^*$ and $\alpha = 1/3$ for $(Ar_4^+)^*$. It follows from Eq. (7) that $I / \ln \overline{N} \sim \overline{N}^{-1/3} W_{rad}$. This means the probability of $(Ar_2)^*$ neutral excimers radiating from one cluster is proportional to the total number of atoms in the cluster ($W_{rad} \sim \overline{N}$), while the emission from charged complexes $(Ar_4^+)^*$ depends on the number of atoms in some near-surface area of the cluster ($W_{rad} \sim \overline{N}^{2/3}$). Indeed, in case of ionization of an argon atom from inner layers of the cluster, the probability of its recombination with the electron before radiating from the $(Ar_4^+)^*$ state is rather high. When an atom from a near-surface layer is ionized, the probability of its recombination is lower since the electron is more likely to leave the cluster.

A different situation is observed for smaller clusters ($\overline{N} \leq 1000$ at/cl). Here, the $I / \ln \overline{N}$ vs. $\overline{N}$ dependences for both bands are linear within the experimental error. It is known from electron diffraction studies that argon clusters of these sizes are quasicrystalline with a 5-fold axis of symmetry (polyicosahedral and multilayer icosahedron structures) [16], while clusters with a few thousand of atoms are characterized by a crystalline fcc structure [17]. If we relate the different behavior of the reduced integrated intensities of the $(Ar_2)^*$ and $(Ar_4^+)^*$ bands to the cluster structure, we can use the cathodoluminescence data to find the cluster-size range in which a passage from icosahedral (multilayer icosahedron) to crystalline (fcc) structures takes place: it can be seen from Fig. 2 that it corresponds to $\overline{N}$ = 1000-1800 at/cl. The deviation of the dependences of reduced intensities of the $(Ar_2)^*$ and $(Ar_4^+)^*$ bands for icosahedral clusters towards lower values with respect to their behavior in the fcc phase can be qualitatively explained by a greater binding energy $\varepsilon_0$ of atoms in an icosahedral cluster [18] (see the factor $\frac{T_*}{\varepsilon_0}$ in Eq. (3)), while the close intensity values for both bands are due to the fact that in small clusters near-surface layers occupy a large fraction of the cluster volume.

**Conclusion**

We propose a new approach to quantitatively analyze integrated intensity $I$ of bands in luminescence spectra of argon in a wide range of average cluster size $\overline{N}$ from 500 to 8900 at/cl.

Our spectroscopic study demonstrates that the fraction of substance $c_{cl}$ condensed into clusters is proportional to the logarithm of their average size, $c_{cl} \sim \ln \overline{N}$.

It is found that in the case of crystalline fcc clusters of argon with $\overline{N} \geq 1800$ at/cl the emission of the vibrationally relaxed neutral excimer molecules $(Ar_2)^*$ comes from the entire cluster ($I / \ln \overline{N} \sim \overline{N}^{2/3}$), while the charged excimer complexes $(Ar_4^+)^*$ radiate from its near-surface layers ($I / \ln \overline{N} \sim \overline{N}^{1/3}$).

The cluster-size range in which a transition between the multilayer icosahedron quasicrystalline structure and the crystalline fcc structure of argon clusters is determined using the cathodoluminescence technique: it corresponds to the average cluster size values $\overline{N}$ = 1000-1800 at/cl.

A preliminary analysis of the data on xenon and krypton allows us to believe that the proposed approach can also be used to analyze luminescence spectra from clusters of other solid rare gases.

The authors would like to thank S.I. Kovalenko, O.G. Danylchenko, and O.P. Konotop for the fruitful discussion of the paper results.

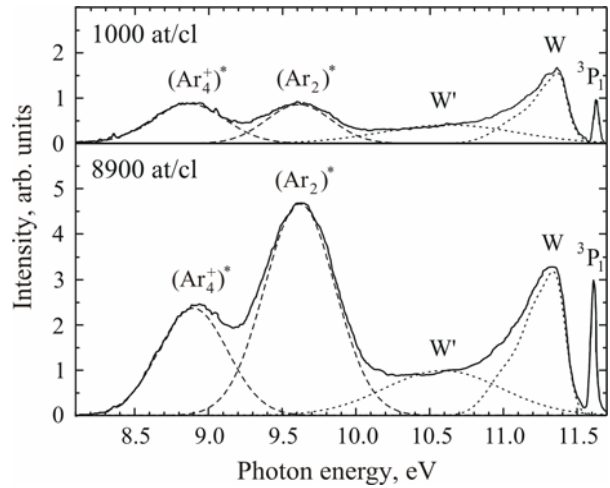

Fig. 1. Luminescence spectra from substrate-free argon clusters with average sizes $\bar{N} \approx 1000$ and 8900 at/cl in the spectral regions of the $(Ar_2)^*$, $(Ar_4^+)^*$, W′, and W bands. Experimental data are shown as solid curves. Decomposition of the spectra into components is shown as dashed ($(Ar_2)^*$ and $(Ar_4^+)^*$) and dotted (W′ and W) curves.

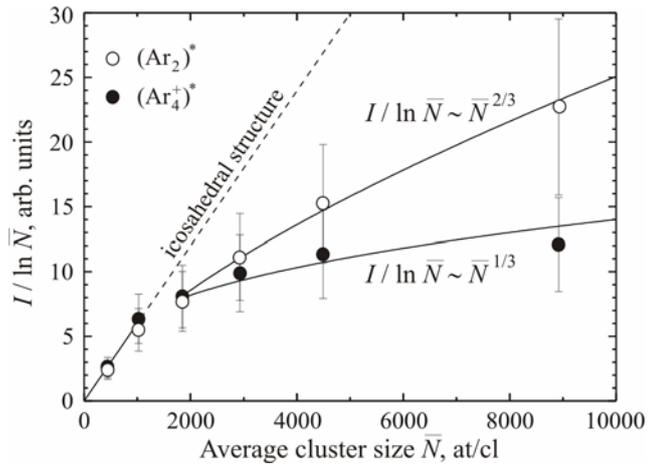

Fig. 2. Average cluster size ($\bar{N}$) dependence of integrated intensity ($I$) reduced by $\ln \bar{N}$ for $(Ar_2)^*$ and $(Ar_4^+)^*$ bands. Regions of icosahedral structure for $\bar{N} \leq 1000$ at/cl, in which $I / \ln \bar{N} \sim \bar{N}$ for both bands, and fcc structure for $\bar{N} \geq 1800$ at/cl, where $I / \ln \bar{N} \sim \bar{N}^{2/3}$ for the $(Ar_2)^*$ band and $I / \ln \bar{N} \sim \bar{N}^{1/3}$ for the $(Ar_4^+)^*$ band, can be seen.